\documentclass[aps,twocolumn,showpacs,amsmath,amssymb,superscriptaddress,prl]{revtex4-1}

\usepackage{graphicx}
\usepackage{dcolumn}
\usepackage{bm}
\usepackage[abs]{overpic}
\usepackage{here}
\usepackage{color}
\usepackage{url}
\usepackage{accents}
\usepackage{subfigure}
\usepackage{ulem}
\def\up{\uparrow}
\def\down{\downarrow }
\def\Vec#1{\bm{#1}}

\DeclareMathOperator{\re}{Re}
\DeclareMathOperator{\im}{Im}

\begin{document}


\title{
DMFT reveals the non-Hermitian topology in heavy-fermion systems
}
\author{Yuki Nagai}
\affiliation{CCSE, Japan  Atomic Energy Agency, 178-4-4, Wakashiba, Kashiwa, Chiba, 277-0871, Japan}
\affiliation{Department of Physics, Massachusetts Institute of Technology, Cambridge, MA 02139, USA}

\author{Yang Qi}
\affiliation{Department of Physics, Fudan University, Shanghai 200433, China}
\affiliation{Department of Physics, Massachusetts Institute of Technology, Cambridge, MA 02139, USA}

\author{Hiroki Isobe}
\affiliation{Department of Physics, Massachusetts Institute of Technology, Cambridge, MA 02139, USA}

\author{Vladyslav Kozii}
\affiliation{Department of Physics, Massachusetts Institute of Technology, Cambridge, MA 02139, USA}
\affiliation{
Department of Physics, University of California, Berkeley, CA 94720, USA
}
\affiliation{
Materials Sciences Division, Lawrence Berkeley National Laboratory, Berkeley, CA 94720, USA
}

\author{Liang Fu}
\affiliation{Department of Physics, Massachusetts Institute of Technology, Cambridge, MA 02139, USA}

\date{\today}

\begin{abstract}
We find that heavy fermion systems can have bulk ``Fermi arcs'', with the use of the non-Hermitian topological theory.
In an interacting electron system,
the microscopic many-body Hamiltonian is Hermitian, but the one-body
quasiparticle Hamiltonian  is non-Hermitian due to the finite quasiparticle lifetime.
We focus on heavy electron systems as a stage of finite lifetime quasiparticles with two lifetimes, since quasiparticle lifetimes for f-electrons and c-electrons should be different.
Two lifetimes induce exceptional points (EPs) of the non-Hermitian quasiparticle Hamiltonian matrix in momentum space.
The line connecting between two EPs characterizes the bulk Fermi arcs.
With the use of the dynamical mean field theory (DMFT) calculation, we confirm our statement in Kondo insulators with a momentum-dependent hybridization in two-dimensions.
We show that the concept of the EPs in the non-Hermitian quasiparticle Hamiltonian is one of powerful tools to predict new phenomena in strongly correlated electron systems.
\end{abstract}

\maketitle
\paragraph{Introduction ---}
A standard theory for metals describes Fermi surfaces that are manifolds with codimension-one [closed loops in two dimensions (2D) and closed surfaces in three dimensions (3D)] in momentum space.
In semimetals, there are codimension-two Fermi nodes and band crossings: nodal points in 2D and nodal lines in 3D, respectively.
However, the pseudogap phase of hole-doped cuprate high-$T_c$ superconductors exhibits Fermi arcs, which are opened Fermi surfaces with endpoints.\cite{Keimer}.

In recent works\cite{Huitao,Vlad,Papaj}, it was revealed that nodal points in 2D Dirac semimetals can become Fermi arcs, the locations of which are determined by a topological theory of finite-lifetime quasiparticles.
In quantum many-body systems, quasiparticles generically have finite lifetimes, resulting from scattering.
Hence, they are described by non-Hermitian Hamiltonians, whose imaginary part reflects the finite lifetimes.
These non-Hermitian matrices can have topological features called the ``exceptional points'' (EPs), where the Hamiltonian becomes non-diagonalizable\cite{Kato}.
It was then shown\cite{Huitao,Vlad,Papaj} that the existence of EPs lead to Fermi arcs, which end on the EPs.
In previous studies, it was shown that two-lifetime models\cite{Vlad}, where two electron bands with finite hybridization have different lifetimes originating from the electron-phonon scattering, often have EPs and thus also Fermi arcs.
Elastic electron-impurity scattering also induces EPs and Fermi arcs\cite{Papaj}.

In this work, we demonstrate how two distinct lifetimes due to electron-electron interaction naturally appear in heavy fermion systems.
Heavy-fermion systems can be described by the periodic Anderson model\cite{Tsunetsugu}.
This model naturally features two lifetimes,
since there are two kinds of electrons, itinerant $c$-electrons and localized $f$-electrons.
The $f$-electrons feel much stronger electron-electron interaction than the $c$-electrons, which leads to different lifetimes of the two bands.

We show that a Kondo semimetal, a Kondo lattice system with a small energy gap such as CeNiSn\cite{Nakamoto}, naturally has Fermi arc.
The Kondo semimetal is understood as a Kondo insulator with a momentum-dependent hybridization gap.
A Kondo insulator exhibits an insulating behavior at low temperatures, since the volume of the Fermi surface counts both $c$- and $f$-electrons and the hybridization of the Fermi surfaces makes this system gapped. However, it exhibits a metallic behavior with a $c$-electron Fermi surface at high temperatures, since the contribution from the $f$-electrons is blurred due to its finite lifetime, which decreases as the temperature raises\cite{PColeman}.
The characteristic crossover temperature between the two behaviors is determined by comparing the magnitude of the hybridization gap and the temperature-dependent inverse lifetime of the $f$-electron.
As we will show, in the Kondo semimetal with a momentum-dependent gap, the crossover happens at different temperatures for different points on the Fermi surface.
Thus, a Kondo semimetal has a Fermi arc, which is a partial Fermi surface ending on EPs.

In this paper, we study a periodic Anderson model with a momentum-dependent hybridization gap in two dimensions.
Using a perturbation theory, we argue that the difference between electron correlations in $c$- and $f$-bands naturally lead to two lifetimes (in fact, under certain assumptions, the self-energy of $c$-electrons vanishes to all orders of perturbations), which results in EPs and Fermi arcs at certain temperatures.
To confirm our statement numerically, we adopt the dynamical mean field theory (DMFT) with the numerically exact continuous-time quantum Monte Carlo solver.
The spectral function calculated by the DMFT clearly shows that there is the Fermi arc defined by the region where  the spectral function has a peak at zero energy.
These Fermi arcs due to the EPs can be observed by
angle-resolved photoemission spectroscopy (ARPES).

\paragraph{Exceptional points of the quasiparticle Hamiltonian ---}
We introduce the EPs and the finite-lifetime quasiparticle Hamiltonian for strongly correlated electron systems\cite{Huitao,Vlad}.

We first introduce quasiparticle Hamiltonian $H$:
\begin{align}
\label{eq:Hq}
H(\Vec{k},\omega) &\equiv H_{0}(\Vec{k}) + \Sigma(\Vec{k},\omega),
\end{align}
which can be extracted from the retarded electron Green's function $G^{R}(\Vec{k},\omega) = (\omega - H(\Vec{k},\omega))^{-1}$,
where $H_{0}(\Vec{k})$ is the single-particle Hamiltonian and $\Sigma(\Vec{k},\omega)$
is the electron's self-energy that includes the effect of electron-electron interaction.
Note that the quasiparticle Hamiltonian $H(\Vec{k},\omega)$ can be {\it non-Hermitian} and that its
spectrum $\mathcal{E}_{n}(\Vec{k},\omega)$ can be complex:
In general, $\Sigma$ is a sum of the Hermitian part $\Sigma'$ and the non-Hermitian part $\Gamma$: $\Sigma = \Sigma'- i \Gamma$.
$\Sigma'$ renormalizes the bare band structure, while $\Gamma$ leads to finite quasiparticle lifetimes,
resulting in the non-Hermitian self-energy and Hamiltonian\cite{Huitao,Vlad}.
The spectrum $\mathcal{E}_{n}(\Vec{k},\omega)$ determines the complex poles $\omega=E_{n}(\Vec{k})$ of the Green's function
according to $E_{n}=\mathcal{E}_{n}(\Vec{k},E_{n})$.
In the vicinity of a first-order pole, the Green's function takes the form $G^{R}(\Vec{k},\omega) \sim \frac{1}{\omega - E_{n}(\Vec{k})}$.
The real part of $E_{n}(\Vec{k})$ determines the quasiparticle's dispersion, while its imaginary part defines the quasiparticle's inverse lifetime.

The important difference between Hermitian and non-Hermitian Hamiltonians is that the latter
can be non-diagonalizable at certain momenta.
Those points are called ``exceptional points'' in the mathematical physics literature\cite{Kato}.
At the EPs, linearly independent eigenstates do not span the full Hilbert space.
One of the authors showed that these EPs in two and higher dimensions are topologically stable\cite{Huitao}.

Topological exceptional points can appear in the quasiparticle spectrum of zero- and small-gap materials
such as Dirac semimetals in two dimensions\cite{Vlad}.
We show that this scenario is also applicable to heavy Fermion systems, where the two orbitals correspond to itinerant and localized
electrons and the hybridization gap can have a momentum-dependent form factor\cite{IkedaMiyake,Dzero,Zhong}.

 \begin{figure*}[t]
\begin{center}
    \begin{tabular}{p{2 \columnwidth}} 
     \resizebox{2 \columnwidth}{!}{\includegraphics{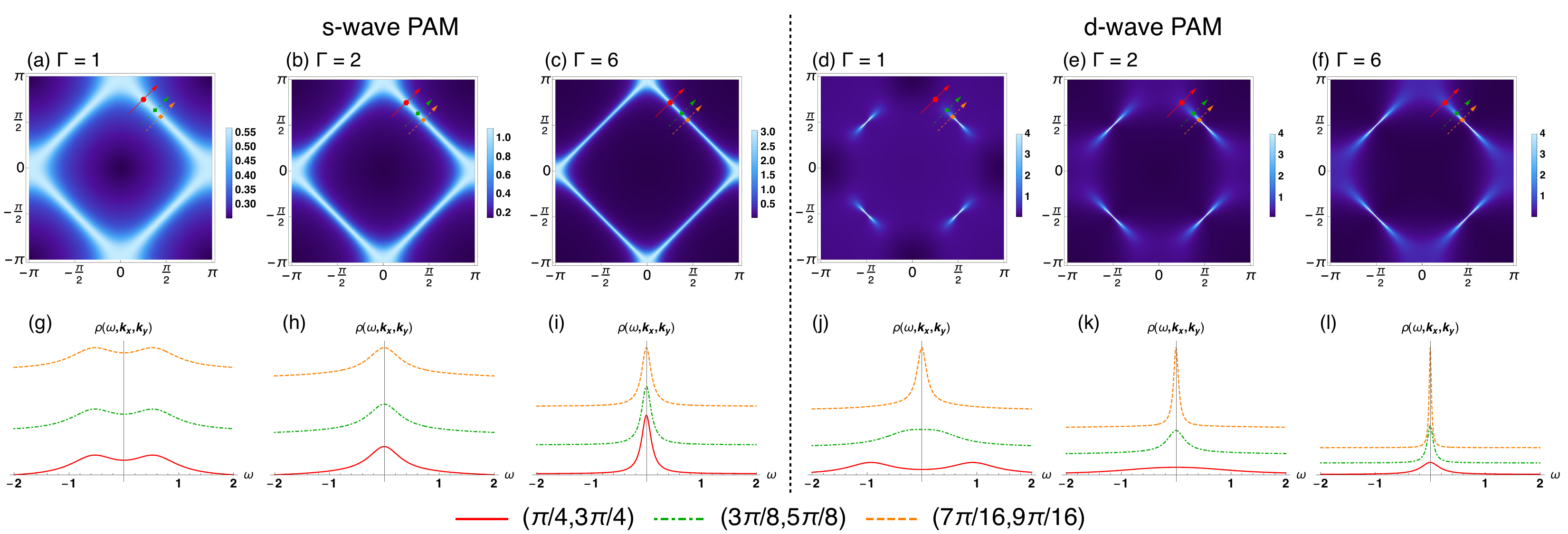}}
   \end{tabular}
\end{center}
\caption{
\label{fig:fig1}
(Color online)
(a)-(f): Momentum dependence of the zero-energy spectral functions $\rho(\omega=0+i\eta,k_{x},k_{y})$ with $\eta = 0.01$. We set $v = 0.75$, $t=1$, and $t_f = 0$.
The red circle, green square, and orange diamond denote points at $(k_{x},k_{y}) = (\pi/4,3\pi/4)$, $(3\pi/8,5\pi/8)$, and $(7\pi/16,9\pi/16)$, respectively.
(g)-(l): Energy dependence of the spectral functions $\rho(\omega,k_{x},k_{y})$ at these points in momentum space.
The results for the $s$-wave periodic Anderson model (PAM) are shown in (a)-(c) and (g)-(i).
The results for the $d$-wave PAM are shown in (d)-(f) and (j)-(l).
The $f$-electron inverse lifetime $\Gamma$ is fixed to $\Gamma = 1$ for (a)(g)(d) and (e), $\Gamma = 2$ for (b)(h)(e) and (k),  and $\Gamma = 6$ for (c)(i)(f) and (l).}
\end{figure*}

\paragraph{Self-energy in heavy fermion systems ---}
Let us now demonstrate how two different lifetimes naturally occur in heavy fermion systems.
We discuss how the self-energy leads to EPs and Fermi arcs, when combined with a momentum-dependent hybridization gap.
We consider a periodic Anderson model (PAM) defined as
\begin{align}
\label{eq:H}
{\cal H} &= \sum_{\Vec{k} \sigma \sigma'}
\left(\begin{array}{cc}
f_{\Vec{k} \sigma}^{\dagger} & c_{\Vec{k} \sigma}^{\dagger} \end{array}\right)
[H_{0}(\Vec{k})]_{\sigma \sigma'}
\left(\begin{array}{c}
f_{\Vec{k} \sigma'} \\
c_{\Vec{k} \sigma'} \end{array}\right)
+ {\cal H}_{int},
\end{align}
where
\begin{align}
  \label{eq:H0}
[H_{0}(\Vec{k})]_{\sigma \sigma'}&= \delta_{\sigma \sigma'}  \left(\begin{array}{cc} \epsilon_{f\Vec{k}} &
v_{\Vec{k}}
 \\
v_{\Vec{k}}
& \epsilon_{c\Vec{k}}
 \end{array}\right),
\end{align}
and ${\cal H}_{int} = U \sum_{i} f_{i \up}^{\dagger} f_{i \up} f_{i \down}^{\dagger} f_{i \down}$.
Here, $c_{\Vec{k} \sigma}$ ($f_{\Vec{k} \sigma}$) is an annihilation operator for itinerant $c$- (localized $f$-) electron with momentum $\Vec{k}$ and spin $\sigma$, with dispersions $\epsilon_{c{\Vec k}}=-2t(\cos k_x+\cos k_y)-\mu$ and $\epsilon_{f\Vec k}=2t_f(\cos k_x+\cos k_y) + \epsilon_{f0} -\mu $, respectively.
We use $t$ as the unit of the energy scale hereafter, and $\mu$ here is the chemical potential.
We consider the case of half filling, which corresponds to $\mu = 0$ and $\epsilon_{f0}=-U/2$, and is characterized by the particle-hole symmetry. At half filling, the term $\epsilon_{f0}=-U/2$ is exactly cancelled by the Hartree part of the electron's self-energy, so we omit it for the rest of the paper for brevity.
The hybridization gap is expressed as $v_{\Vec{k}} = v $ for $s$-wave case which describes the Kondo insulators and $v_{\Vec{k}} = v(\cos k_{x} - \cos k_{y}) $ for $d$-wave case which describes Kondo semimetals, respectively. In this work, we focus on the $d$-wave gap, which has point nodes on the Fermi surface at $(k_x,k_y) = (\pm \pi/2, \pm \pi/2)$~\cite{sup}.

It is known that the PAM exhibits a number of strongly correlated phenomena showing the non-Fermi-liquid behavior \cite{Amaricci2008} and undergoing the Mott transition \cite{Sordi2007,Amaricci2017,Amaricci2012} near some particular fillings. 
In our work, we stay away from these special fillings, i.e., in the regime where the Fermi liquid description remains adequate.

Due to the fact that the $c$ electrons in model \eqref{eq:H} are not interacting, only $f$ electrons have a finite lifetime:
This implies the following form of
\begin{equation}
  \label{eq:seff_main}
  [\Sigma(\Vec{k},\omega)]_{\sigma \sigma'} = \begin{pmatrix}
    \Sigma^{f}(\Vec{k},\omega) \delta_{\sigma \sigma'} & 0 \\ 0 & 0
  \end{pmatrix}.
\end{equation}
This point is explained in Sec.~S2. in the supplemental material.
Thus, the two-lifetime description naturally arises in the periodic Anderson model.

The $f$-electron self-energy is computed using a second-order perturbation theory in Sec. S2 of the supplemental material (see also, Ref.~\cite{Yamada1986}).
At temperatures comparable or higher than the hybridization scale $v$, we can ignore the momentum dependence of $\Sigma(\omega)$. This is because the finite-temperature broadening renders the $f$-band approximately flat.
Hence, the onsite interaction and a flat band together result in a local self-energy.
Since we are interested in the spectrum of low-energy electron excitations, we focus on low-frequency self energies.
In particular, in the zero-frequency limit, $\Sigma^f$ can be treated as a purely imaginary constant, $\Sigma^f(\bm k, 0)=-i\Gamma$,
where the real part vanishes because the particle-hole symmetry pins the energy of the $f$-electron at zero.
Furthermore, at small frequencies, one can expand $\Sigma^f$ as the following:
\begin{equation}
\label{eq:Sf-w}
\Sigma^f(\omega)=-a_1\omega-i(\Gamma+a_2\omega^2)+\cdots.
\end{equation}
Our DMFT calculations presented later show that coefficients $a_1$ and $a_2$ are generally positive.
The inclusion of the positive higher-order terms $a_1$ and $a_2$ will not change the qualitative features of the electron spectrum function discussed next.
In particular, the effective of including $a_1$, which is the leading term of the real part of $\Sigma^f$, is a reduction of the widths of some bands: it renormalizes $t_f$ to $Z t_f$, $\Gamma$ to ${Z} \Gamma$, and $v_k$ to $Z^{1/2} v_k$, respectively, while keeping $t_c$ unaffected. (This is analyzed in details in Sec.~S2 of the SM.)
Here, $Z \equiv (1 + a_1)^{-1}$.

\paragraph{Bulk Fermi arcs ---}
We now discuss the electron spectral function, and show the Fermi arcs in the $d$-wave PAM.
We plot the electron spectral function from \eqref{eq:Hq}.
The result is shown in Fig.~\ref{fig:fig1}.

In general, the inverse lifetime $\Gamma$ is small at low temperatures and large at high temperatures, reflecting a long and a short quasiparticle lifetime, respectively.
Hence, we demonstrate low-temperature and high-temperature phenomena using $\Gamma=1$, $\Gamma = 2$ and $\Gamma=6$, respectively.
In the $s$-wave PAM, there is a gap with $\Gamma = 1$ in the whole momentum space as shown in Figs.\ref{fig:fig1}(a) and (g).
With increasing $\Gamma$, the gap is closed on a Fermi surface.
This is the standard crossover behavior between a Kondo insulator at a low temperature and a small $c$-electron Fermi surface at a high temperature.
In the $d$-wave PAM, on the other hand, we can clearly see the bulk Fermi arcs at the low temperature, as shown in Figs.\ref{fig:fig1}(d) and (j). The Fermi arcs grow into a full $c$-electron Fermi surface at the high temperature.

The Fermi arcs observed in the $d$-wave PAM can be explained as follows.
With $\Sigma=-i\Gamma -a_1\omega$,
the complex poles of the Green's function are given by
\begin{align}
E_{\pm}(\Vec{k}) =&
\pm \frac{Z}{2} \sqrt{(M_{\Vec{k}}-i \Gamma)^{2} + 4 Z^{-1}|v_{\Vec{k}}|^{2}} \nonumber \\
 &+\frac{Z}{2}(\epsilon_{f\Vec k}-i \Gamma+Z^{-1}\epsilon_{c\Vec k} ),\label{eq:Epm}
 \end{align}
with $M_{\Vec{k}} \equiv \epsilon_{f\Vec k}-Z^{-1}\epsilon_{c\Vec k}$.
This complex spectrum is similar to that for the Dirac semimetals in two dimensions if we replace $M_{\Vec k}$
and $|v_{\Vec{k}}|$ with
$v_{x} k_{x}$ and $v_{y} k_{y}$, respectively (and assuming $Z=1$)\cite{Vlad}.

The spectral function $\rho(\omega,\Vec{k}) = -\frac1{\pi}\text{Im Tr } (G^R - G^A)$ with $G^A \equiv \left[  G^R \right]^\dagger$ is given by
\begin{align}
\rho(\omega,\Vec{k}) = - \frac{2}{\pi} {\rm Im} \: \left[\frac{A_{\Vec{k}}}{\omega - E_{+}(\Vec{k})}+\frac{B_{\Vec{k}}}{\omega - E_{-}(\Vec{k})} \right],  \label{eq:spec}
\end{align}
with
\begin{align}
A_{\Vec{k}} = \frac{(1+Z) E_+(\Vec{k}) - Z (\epsilon_{c \Vec{k}} + \epsilon_{f \Vec{k}} - i \Gamma)}{E_+(\Vec{k}) - E_-(\Vec{k})} \nonumber \\ B_{\Vec{k}} = \frac{Z (\epsilon_{c \Vec{k}} + \epsilon_{f \Vec{k}} - i \Gamma) - (1+Z) E_-(\Vec{k}) }{E_+(\Vec{k}) - E_-(\Vec{k})}.
\end{align}
The form-factors $A_{\Vec{k}}$ and $B_{\Vec{k}}$ originate from the non-zero real part of self-energy, and both equal to 1 in case when $a_1 = 0$.

 We see that the spectral function is given by the sum of two Lorentzians, whose locations and widths are given by the real and imaginary parts of $E_\pm$, respectively. In particular, we focus on the line $\epsilon_{c\Vec k}=Z \epsilon_{f\Vec k}$, which is the location of the high-temperature $c$-electron Fermi surface.
Along this line, Eq.~\eqref{eq:Epm} is simplified to the following form,
\begin{equation}
 \label{eq:Epm2}
 E_{\pm}(\Vec{k}) =
 \pm\frac{Z}{2}\sqrt{-\Gamma^2 + 4Z^{-1}|v_{\Vec{k}}|^{2}}+Z \epsilon_{f\Vec k}-i Z \frac{\Gamma}{2}.
\end{equation}

We first focus at a fixed momentum $k$, and discuss how a temperature-dependent $\Gamma$ affects the electron spectrum function.
For simplicity, we consider $Z = 1$($a_1 = 0$).
In general, we expect $\Gamma$ to increase with temperature.
In particular, it should approach zero in the $T=0$ limit.
The competition between the temperature-dependent $\Gamma$ and the hybridization $v_k$ can explain the Kondo crossover.
At low temperatures, when $4 |v_{\Vec k}|^2 > \Gamma^2$, $E_\pm$ have different real parts, and the spectrum is a superposition of two peaks.
The distance between the two peaks, $\sqrt{4|v_{\Vec k}|^2-\Gamma^2}$, represents a hybridization gap.
This indicates that the $f$ and $c$ electrons are hybridized, and the $f$ electrons join the $c$ electrons to contribute to the volume of the Fermi surface.
At high temperature, when $4|v_{\Vec k}|^2 < \Gamma^2$, $E_\pm$ have the same real part, and the spectrum exhibits a single peak located at $\omega=0$ if $\epsilon_{c\Vec k}=\epsilon_{f\Vec k}$. This indicates that there is no gap, and $k$ is on a Fermi surface.
This also means that the $f$-electrons form local moments, which are detached from the Fermi surface formed solely by the $c$ electron.

For the $s$-wave PAM, the crossover between the two behaviors happens at the same temperature where $\Gamma=2v$ for all $\bm k$, and a complete $c$-electron Fermi surface emerges from a Kondo insulator as the temperature raises.
For the $d$-wave PAM, however, at different momenta on the Fermi surface, the crossover happens at different temperatures determined by $v(\Vec k)$.
In particular, for $\Gamma < 2v$, the Fermi surface is separated into two regions by EPs located at $2|v_{\Vec k}| = \Gamma$, where $E_\pm$ become degenerate.
The hybridization gap only appears on part of the crossing lines, where $4|v_{\Vec k}|^2>\Gamma^2$.
The rest of the crossing lines, where $4|v_{\Vec k}|^2<\Gamma^2$, then forms Fermi arcs.
Hence, the arcs are longer with bigger $\Gamma$.
At zero temperature, the model is a Kondo semimetal with point nodes.
As temperature raises and $\Gamma$ increases, the Fermi arcs grow longer, until the full $c$-electron Fermi surface emerges at $\Gamma=2v$.

We notice that, phenomenologically, when two peaks in the spectrum function are closer than their peak widths, they become inseparable experimentally.
Hence, the exact locations of the ending points of the Fermi arcs may differ from the EPs, and they depend on the details of data analysis.
However, the existence of the Fermi arcs is tied to EPs in the quasiparticle Hamiltonian.

\begin{figure}[t]
\begin{center}
    \begin{tabular}{p{0.9 \columnwidth}} 
     \resizebox{0.9 \columnwidth}{!}{\includegraphics{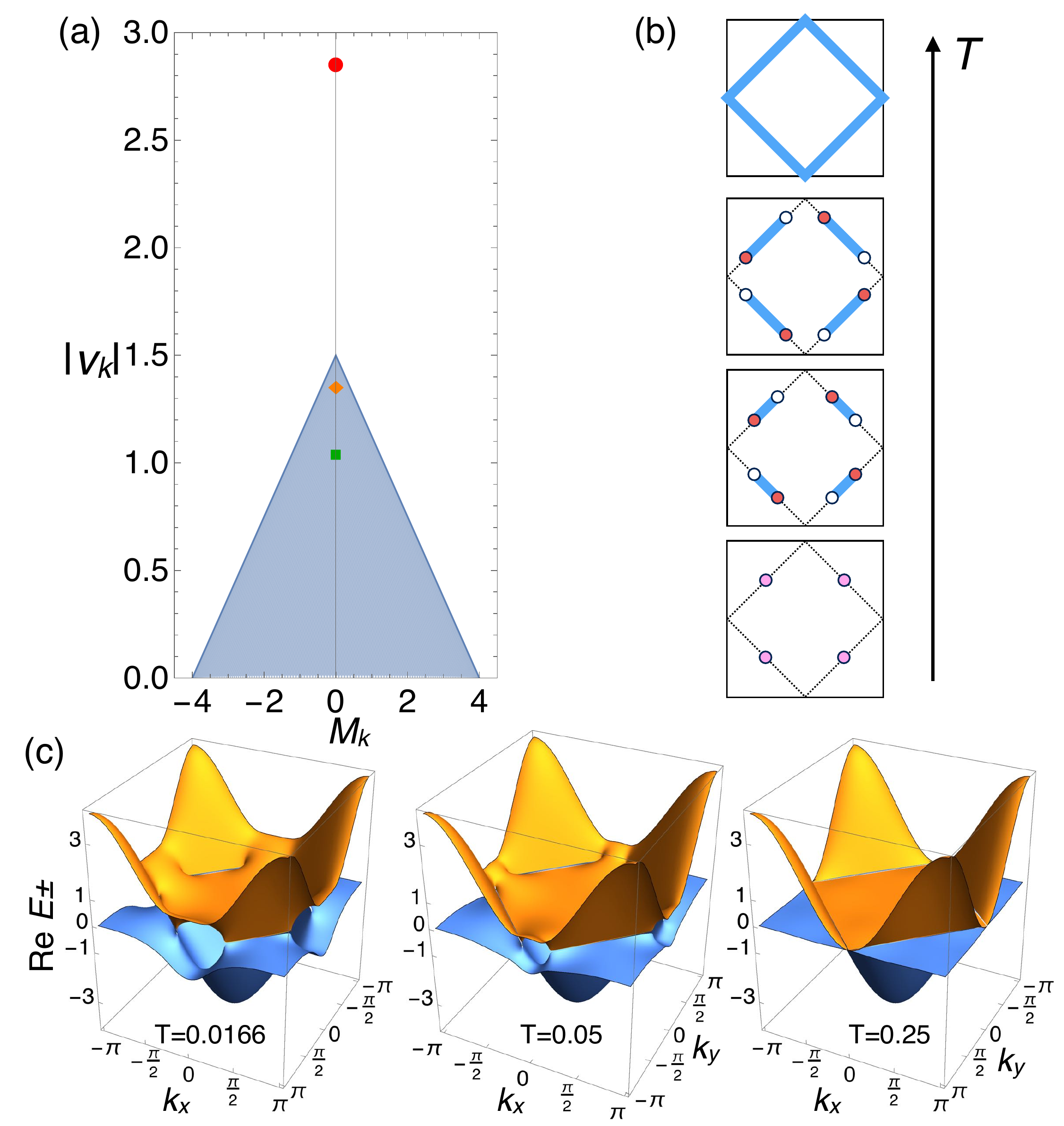}}
   \end{tabular}
\end{center}
\caption{
\label{fig:fig4}
(Color online) (a): Exceptional points (EPs) in $(M_{\Vec{k}}, |v_{\Vec{k}}|)$ space in the two-dimensional model for Kondo insulator.  The shaded region indicates possible $(M_{\Vec{k}}, |v_{\Vec{k}}|)$ region.
The red circle, orange diamond, and green square denote the value of the lifetime at $T=1/4,1/20,1/60$, respectively.
(b): Schematic figures of the EPs. the EPs move from the middle of the line $\epsilon_{c\Vec{k}}=0$ to the corner of the square.
(c): Momentum dependence of the real part of the complex energy spectrum of the non-Hermitian quasiparticle Hamiltonian with different temperatures from $T=1/60$ to $T=1/4$.
}
\end{figure}

\paragraph{Dynamical Mean Field Theory ---}
To treat the heavy fermion systems more accurately, we use the dynamical mean field theory (DMFT).
In the DMFT, we assume that the self-energy does not depend on the momentum:
\begin{align}
\Sigma^{f}(\Vec{k},\omega) = \Sigma^{f}(\omega).
\end{align}
This assumption is appropriate when the temperature is not too low comparison with the maximum of hybridization energy\cite{sup}.
The momentum dependence does not affect the appearance of the bulk Fermi arcs, qualitatively.
We set $v = 0.75$ and $t_f = 0$.
To calculate the local and momentum-dependent spectral functions $\rho(\omega)$ and $\rho(\omega,\Vec{k})$, we utilize the DMFT with the numerically exact segment-based hybridization-expansion
continuous-time quantum Monte Carlo impurity solver (ct-HYB) \cite{Werner}.
We point out that there is no fermion sign problem because of the off-diagonal elements of the self-energy in the spin space
are zero in our two-orbital system\cite{NagaiDMFT}.
After confirming that the self-energy for $c$ electrons is zero in the two-orbital DMFT calculation, we use one-orbital DMFT calculation with integrating out the $c$-electron degree of freedom.
Here, the effective impurity problem is solved by an open-source program package, $i$QIST\cite{iqist}.
We set $U = 8$.

 \begin{figure*}[t]
\begin{center}
    \begin{tabular}{p{1.5 \columnwidth}} 
     \resizebox{1.5 \columnwidth}{!}{\includegraphics{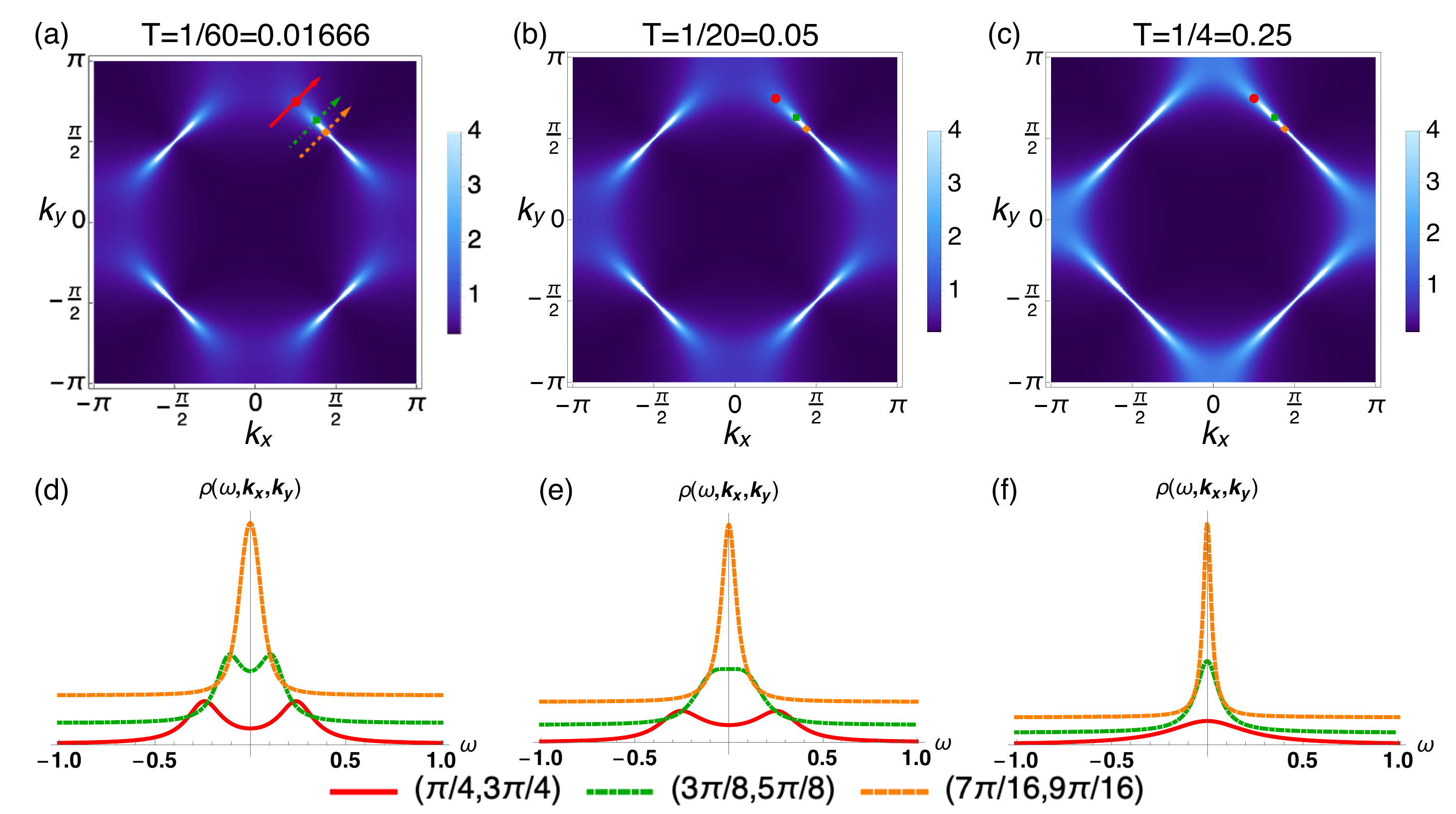}}
   \end{tabular}
\end{center}
\caption{
\label{fig:fig2}
(Color online)
(a)-(c): Momentum dependence of the zero-energy spectral functions $\rho(\omega=0,k_{x},k_{y})$ in the DMFT calculation.
The red circle, green square, and orange diamond denote points at $(k_{x},k_{y}) = (\pi/4,3\pi/4)$, $(3\pi/8,5\pi/8)$, and $(7\pi/16,9\pi/16)$, respectively.
(d)-(f): Energy dependence of the spectral functions $\rho(\omega,k_{x},k_{y})$ on these points in momentum space.
Temperatures are (a)(d) $T=1/60$, (b)(e) $T=1/20$ and (c)(f) $T=1/4$, respectively. For these temperatures, the fitting of
the DMFT data gives us $\Gamma=2.06,$ 2.71, and 5.71, correspondingly. We set $t=1$, $v=0.75$, $t_f=0$ and $U=8$.
}
\end{figure*}

Figure \ref{fig:fig2} shows the spectral functions and the Fermi arcs.
To calculate the real-frequency self-energy $\Sigma(\omega)$ and the spectral function $\rho(\omega,\Vec{k})$,
we use the Pad\'{e} approximation as the method for the numerical analytic continuation of the self-energy $\Sigma(i \omega_{n} \rightarrow \omega)$.
After the Pad\'{e} approximation, we fit the self-energy with the form $\Sigma_{f}(\omega) = -(a_{1}\omega+a_{3}\omega^{3})-i(\Gamma+a_{2}\omega^{2}+a_{4}\omega^{4})$ in the region $-1 < \omega < 1$.
We find that the set of parameters $(\Gamma,a_{1},a_{2},a_{3},a_{4})$ equals
(5.70895, 2.28419, 1.01368, 1.15838, 0.580893) at $T=1/4$,
(2.70929, 9.54693, 5.49701, -2.19662, 5.38239) at $T=1/20$, and
(2.06142, 11.8766, 16.49, -10.2929, -6.21199) at $T=1/60$, respectively.
Along the Fermi line without hybridization defined by $M_{\Vec{k}}=0$ close to $\Vec{k} = (0,\pm \pi)$ and $(\pm \pi,0)$, the spectral function $\rho(\omega,\Vec{k})$ begins to have a dip at $\omega=0$ by decreasing temperature, as shown in Fig.~\ref{fig:fig2}(d)-(f).
We confirm that the spectral functions as a function of $\omega$ and $\Vec{k}$ show that
there is a gap at zero energy in the low temperature region.
These results indicate the existence of the Fermi arc, since the line where the spectral function has a peak at the zero energy has boundaries.

\paragraph{Summary ---}
In this work, we studied heavy-electron systems using PAMs.
Although the microscopic many-body Hamiltonian is Hermitian, the one-body
quasiparticle Hamiltonian is non-Hermitian because of finite quasiparticle lifetimes, which is due to electron interactions.
In particular, we demonstrate that the difference in electron correlations between $c$- and $f$-electrons in PAMs naturally leads to two lifetimes in the non-Hermitian quasiparticle Hamiltonian.
This, combined with a $d$-wave hybridization gap, leads to the existence of EPs in the Hamiltonian and ``Fermi arcs'' in electron spectral function, at suitable temperatures.
Our DMFT calculation with a numerically exact continuous-time quantum Monte Carlo solver confirmed our statement.
These unusual Fermi arcs due to EPs can be observed by ARPES, which measures electron spectral function.
Our discussion is applicable to other systems, such as two- and three-dimensional Kondo insulators with $p$-wave hybridization.
In a two-dimensional system with $p$-wave hybridization ($V_{\Vec{k}}=\Vec{v}_{\Vec{k}} \cdot \Vec{\sigma}$ with $\Vec{v}_{\Vec{k}}=(\sin k_{x},\sin k_{y})$), cross-shaped Fermi arcs appear around $\Vec{k} = (\pm \pi,0)$ and $(0,\pm \pi)$.
In a three-dimensional system, the Fermi arc becomes the three-dimensional ``Fermi arc'' defined by the surface with the boundaries in momentum space,
which will be shown elsewhere.

\paragraph{Acknowledgment ---}
Y.N. would like to acknowledge H. Shen for helpful discussions and comments.
The calculations were performed by the supercomputing system SGI ICE X at the Japan Atomic Energy Agency.
YN was partially supported by JSPS KAKENHI Grant Number 15K00178 and 18K03552, the ''Topological Materials Science'' (No. JP16H00995 and No. 18H04228) KAKENHI on Innovative Areas from JSPS of Japan.
V. K. was partially supported by the Quantum Materials program at LBNL, funded by the US Department of Energy under Contract No. DE-AC02-05CH11231.

Y.N. and Y.Q. contributed equally to this work.

\paragraph{Note added ---} Recently, we learned about the related works \cite{Yoshida} and \cite{Michishita}

\clearpage
\newpage
\widetext
\onecolumngrid

\setcounter{equation}{0}
\renewcommand{\thefigure}{S\arabic{figure}}

\setcounter{figure}{0}

\renewcommand{\thesection}{S\arabic{section}.}
\renewcommand{\theequation}{S\arabic{equation}}
\renewcommand{\thetable}{S\arabic{table}}
\begin{flushleft}
{\Large {\bf Supplemental material for ``DMFT reveals the non-Hermitian topology in heavy-fermion systems''}}
\end{flushleft}

\begin{flushleft}
{\bf S1. Density of states from the vicinity of a Dirac node}
\end{flushleft}

The energy dispersion near a node located at $(\pm\pi/2,\pm\pi/2)$ is described by the effective Hamiltonian to the first order in $k$:
\begin{align}
H_0(\bm{k}) &= v_x k_x \sigma_z + v_y k_y \sigma_x - w k_x \sigma_0 \nonumber\\
&= \begin{pmatrix}
(v_x-w) k_x & v_y k_y \\
v_y k_y & -(v_x+w) k_x
\end{pmatrix}
.
\end{align}
The axes are rotated by $45^\circ$ for convenience and we find the correspondence $v_x = \sqrt{2}(t_c + t_f)$, $w = -\sqrt{2}(t_c - t_f)$, and $v_y = \sqrt{2}v$. It can be interpreted as follows: The $k_x$ direction is chosen to be perpendicular to the Fermi surface, along which hybridization is absent.  It corresponds to the $\Gamma$--$M$ direction.  Then, $k_y$ is parallel to the Fermi surface and it describes the hybridization of the two bands.  The two different bands allows the parameter $w$, describing the tilt of the Dirac cone.  Along the $k_x$ direction, the slower velocity $v_x-w$ of the first component corresponds to the $f$ band, and the faster $v_x+w$ to the $c$ band.  The off-diagonal terms represent hybridization of the two bands.

The local density of states $\rho(\omega)$ is obtained by
\begin{equation}
\rho(\omega) = -\frac{1}{\pi} \operatorname{Im} \int\frac{d^2k}{(2\pi)^2} G^R(\bm{k},\omega),
\end{equation}
with the retarded Green's function
\begin{align}
G^R(\bm{k},\omega) &= \left[\omega+i\delta -H_0(\bm{k}) -\Sigma^R(\omega)\right]^{-1}.
\end{align}
Here we assume that the self-energy $\Sigma^R(\omega)$ takes the form
\begin{equation}
\Sigma^R(\omega) = \Sigma^R_0(\omega)\sigma_0 - \Sigma^R_z(\omega)\sigma_z.
\end{equation}
Without a trace of $2\times 2$ matrices, we can extract the contribution from each band.
When the self-energy $\Sigma^R(\omega)$ is independent of momentum $\bm{k}$, we can evaluate the momentum integration to obtain
\begin{equation}
\rho(\omega) = \frac{1}{4\pi^2 v_x v_y \left(1-\dfrac{w^2}{v_x^2}\right)^{3/2}} \left( \sigma_0 + \frac{w}{v_x} \sigma_z \right) \operatorname{Im} \left( \tilde{\omega} \ln \frac{\Lambda^2}{-\tilde{\omega}^2} \right),
\end{equation}
where $\Lambda$ is the high-energy cutoff and $\tilde{\omega}$ is defined by
\begin{equation}
\tilde{\omega} = \omega + i\delta -\Sigma_0^R(\omega) -\frac{w}{v_x} \Sigma_z^R(\omega).
\end{equation}
We note that the expression of $\rho(\omega)$ is valid for $w^2<v_x^2$ and $\operatorname{Im}\Sigma_0>\operatorname{Im}\Sigma_z$.

When the self-energy $\Sigma$ vanishes, we find an expression for free Dirac fermions, i.e., $\rho(\omega)\propto \omega$.  In contrast, a finite imaginary part of the self-energy $\operatorname{Im}\Sigma(0)$ makes the density of states finite at $\omega=0$: $\rho(0) \neq 0$.

\begin{flushleft}
{\bf S2. The effect of $a_1$ on the dispersion}
\end{flushleft}

In this section, we study the effect of $a_1$, which is the coefficient of the linear-$\omega$ term in the real part of the $f$-electron self-energy [see Eq.~\eqref{eq:Sf-w} in the main text].
We argue that the effect of including $a_1$ can be absorbed into renormalizing the parameters $t_c$, $t_f$, $\Gamma$ and $v_k$ in electron's Green's function.
To see this, we consider the Green's function determined by including the $f$-electron self-energy $\Sigma^f=-i\Gamma-a_1\omega$,
\begin{equation}
  \label{eq:Gf-Ga1}
  G^{-1}=\begin{pmatrix}
    \omega-\epsilon_{f\bm k}+i\Gamma+a_1\omega & v_k\\
    v_k&\omega-\epsilon_{c\bm k}
\end{pmatrix}.
\end{equation}
As discussed in the main text, the electron spectral function can be decomposed into Lorentzian peaks, which are determined by poles of $G_f$ [see Eq.~\eqref{eq:spec}].
The poles of $G_f$ are the solutions of $\det G^{-1}=0$, which can be rewritten as the following form,
\begin{equation}
  \label{eq:Gf-Ga1r}
  \det\begin{pmatrix}
    \omega-\epsilon_{f\bm k}/(1+a_1)+i\Gamma/(1+a_1)& v_k/\sqrt{1+a_1}\\
    v_k/\sqrt{1+a_1}&\omega-\epsilon_{c\bm k}.
\end{pmatrix}=0.
\end{equation}
Hence, we see that $\epsilon_{f\bm k}$, $\Gamma$ and $v_k$ are normalized to $\epsilon_{f\bm k}/(1+a_1)$, $\Gamma/(1+a_1)$ and $v_k/\sqrt{1+a_1}$, respectively.

\begin{figure}
  \subfigure[\label{fig:d3d:arc}]{\includegraphics[width=.45\textwidth]{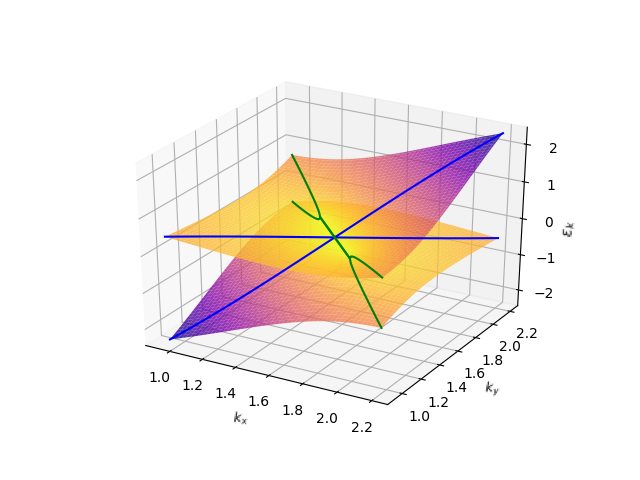}}
  \subfigure[\label{fig:d3d:arc2}]{\includegraphics[width=.45\textwidth]{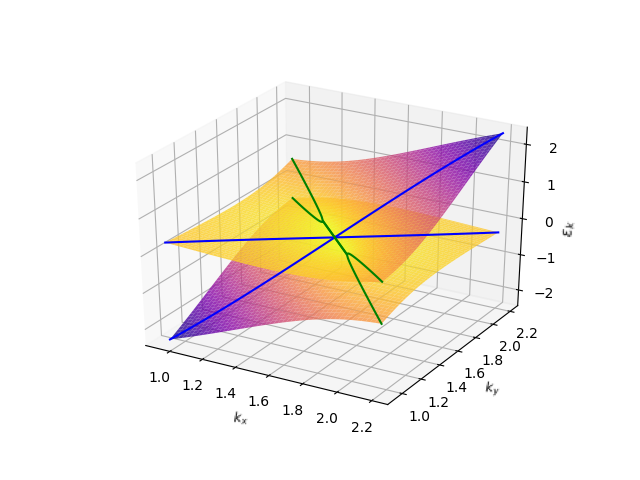}}\\
  \subfigure[\label{fig:d3d:comp1}]{\includegraphics[width=.45\textwidth]{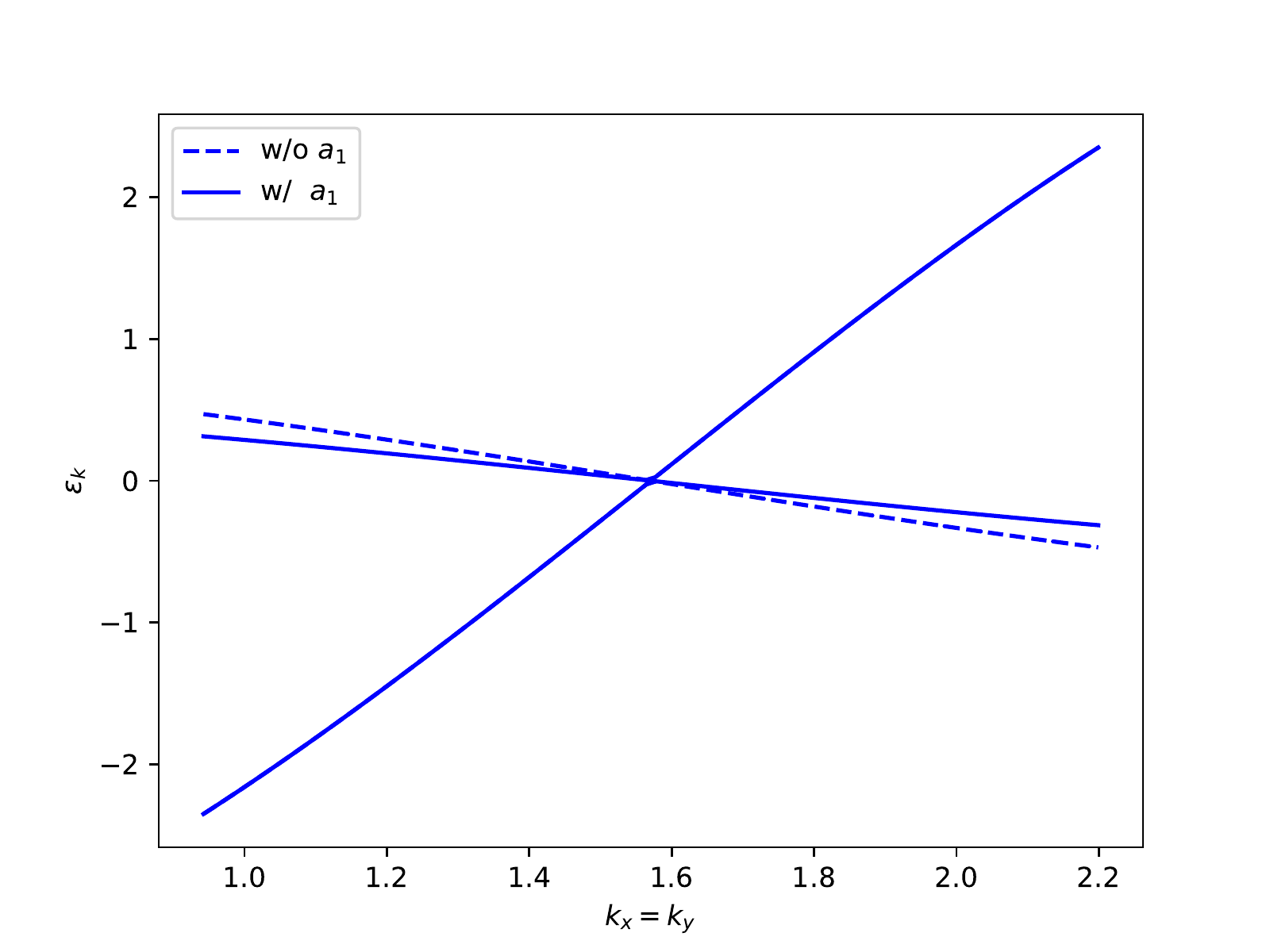}}
  \subfigure[\label{fig:d3d:comp2}]{\includegraphics[width=.45\textwidth]{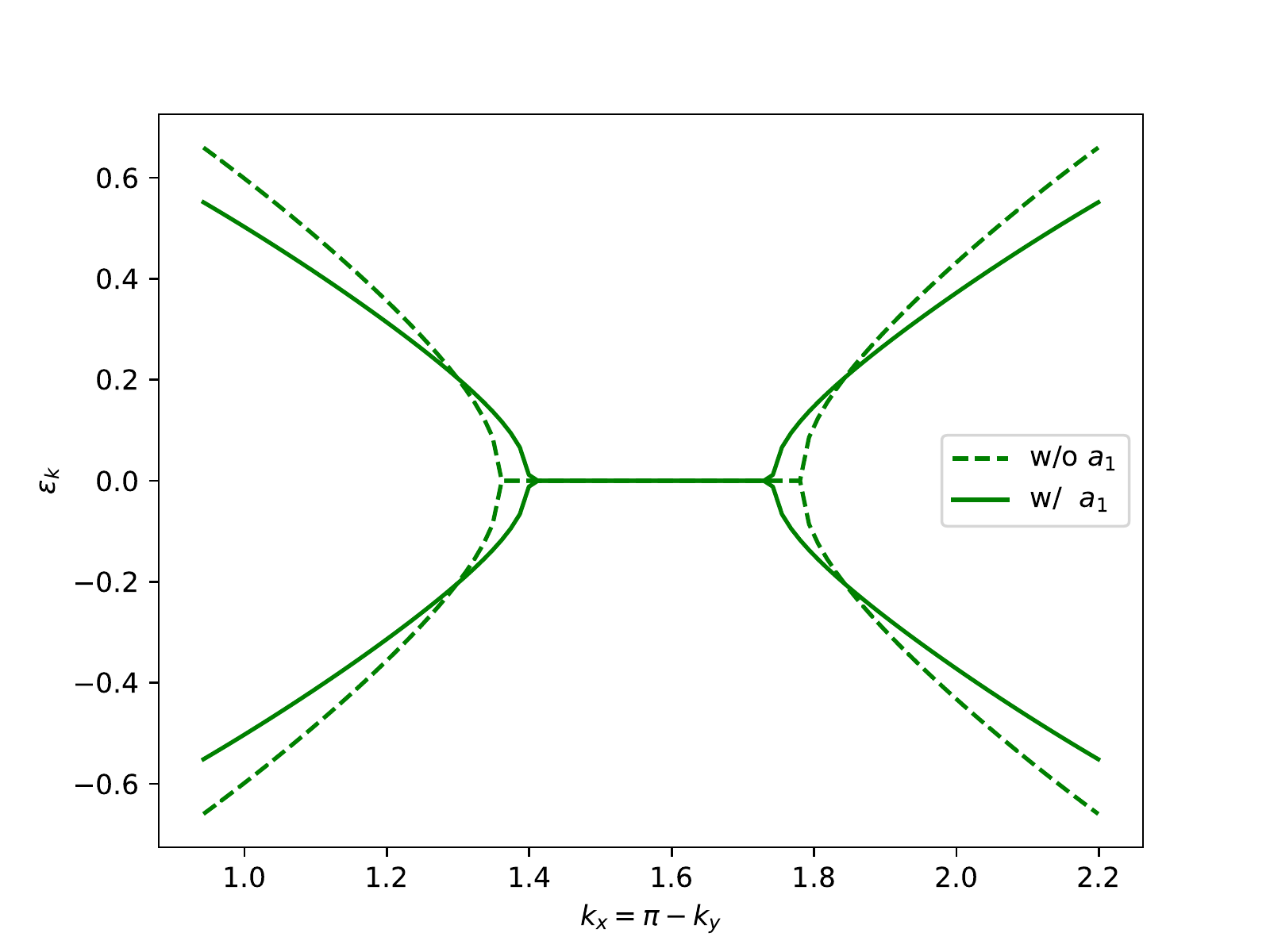}}
  \caption{
  Dispersion (real part of the energy eigenvalues Eq.~(6) of the main
text) near the nodal point. (a) Dispersion near the nodal point without including $a_1$. The colored surface shows the energy of the quasiparticle peaks at different momenta. The blue and green curves highlights the dispersion along the $k_x=k_y$ and $k_x+k_y=\pi$ directions, respectively. The portion where the two green curves collapse together indicates the Fermi arc. (b) The dispersion with $a_1$. The main features remain the same, but the dispersion is renormalized. (c) The dispersion along the $k_x=k_y$ direction. The dashed line and solid line shows the dispersion without and with $a_1$, respectively. The heavy $f$-electron band becomes heavier with $a_1$, but the $c$-electron band is not renormalized. (d) The dispersion along the $k_x+k_y=\pi$ direction. The dashed line and solid line shows the dispersion without and with $a_1$, respectively. 
 After including $a_1$, $v$ becomes smaller, and the Fermi arc, defined by the exceptional points of Eq.~(9) of the main text, becomes shorter.
  All plots are computed with $t=1$, $t_f=0.2$, $v=0.6$, $\Gamma=0.5$, and $a_1=0.5$.}
  \label{fig:d3d}
\end{figure}

The effect of $a_1$ is illustrated in Fig.~\ref{fig:d3d}, which shows
the dispersion of the quasiparticle peaks with and without $a_1$. 
One can see that, after including $a_1$, the bandwidth of $f$-electron and the hybridization gap becomes smaller, and the Fermi arc, defined by Eq.~(9) of the main text, becomes shorter, but the $c$-electron band is not renormalized. 
To see the effect of $a_1$, however, one has to consider the frequency dependence of the spectral function. 
Then, the exceptional points, which are the boundaries of the Fermi arcs, can be determined as points on the Fermi surface where $\rho(\omega)$ starts having a dip at zero frequency instead of a peak, i.e, the zero-frequency peak splits into two peaks at finite frequencies, as shown in Figs.~1 (j)-(l) and 3~(d)-(f) of the main text. At zero frequency, on the other hand, the spectral function does not depend on $Z$ (equivalently, on $a_1$) 
and is given by the following expression:
\begin{align}
\rho(\omega = 0,{\bm k}) &= \frac{1}{\pi} \frac{2 \Gamma (\epsilon_{c{\bm k}}^2 + |v_{\bm k}|^2)}{(\epsilon_{c{\bm k}} \epsilon_{f{\bm k}} -  |v_{\bm k}|^2)^2 +\epsilon_{c{\bm k}}^2 \Gamma^2 }.
\end{align}

\vspace{0.5cm}

\begin{flushleft}
{\bf S3. Self energy from the second-order perturbation theory}
\end{flushleft}

\begin{figure}
  \subfigure[\label{fig:sigma:a}]{\includegraphics{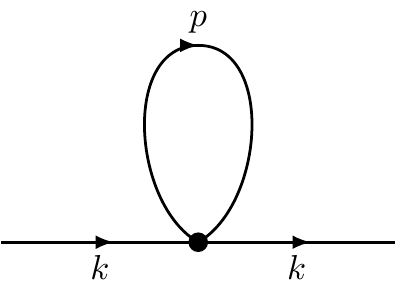}}
  \subfigure[\label{fig:sigma:b}]{\includegraphics{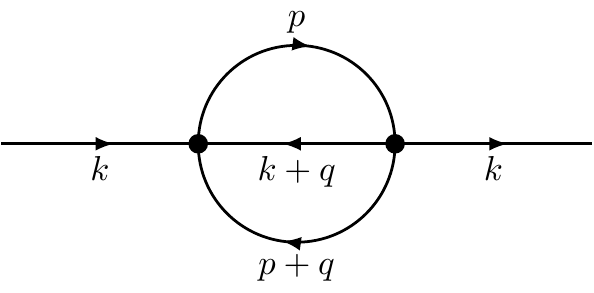}}
  \subfigure[\label{fig:sigma:c}]{\includegraphics{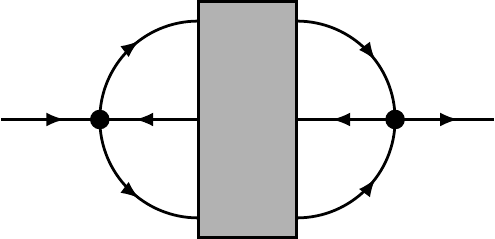}}
  \caption{Feynman diagrams appeared in the perturbation theory of $U$. The solid lines with arrows represent $f$-electron propagators. The black dot represents interacting vertex with coefficient $U$, which has four $f$-electron legs. (a) First-order correction to the fermion self energy. (b) Second-order correction to the fermion self energy. (c) Scheme of high-order diagrams in self-energy correction.}
  \label{fig:sigma}
\end{figure}

In this section, we compute the electron self-energy using a second-order perturbation theory in $U$. In the perturbation theory, the self-energy is given by the one-particle-irreducible (1PI) diagrams. To the second order of $U$, there are two such diagrams, shown in Fig.~\ref{fig:sigma}.

The perturbation theory shows that, since there is no interaction between
$c$-electrons in model \eqref{eq:H}, only $f$-electrons have the self-energy.
To see this, we notice that all fermion propagators in the self-energy diagrams in Fig.~\ref{fig:sigma} are of $f$ fermions. This is because the interaction vertex, represented by the dot in the diagrams, only involves four $f$ electrons, since it comes from the onsite $U$ term in Eq.~\eqref{eq:H}. Therefore, in the matrix of the self-energy, only the diagonal block of the $f$ electron is nonzero:
\begin{equation}
  \label{eq:seff}
  \Sigma_{\alpha\beta} = \begin{pmatrix}
    \Sigma^{ff}_{\alpha\beta} & 0 \\ 0 & 0
  \end{pmatrix}.
\end{equation}
In fact, this block structure even persists to all orders of perturbation. This is because all 1PI diagrams of orders higher than three can be represented by the schematic diagram in Fig.~\ref{fig:sigma}(c), where the self-energy bloc begins and ends with two $U$ vertices. Hence, the incoming and outgoing legs must be of $f$ electrons. Since the imaginary part of the non-Hermitian Hamiltonian in Eq.~\eqref{eq:Hq} comes solely from the imaginary part of the self-energy, the imaginary part also have this block structure: $\Gamma^{cc}=\Gamma^{cf}=\Gamma^{fc}=0$. 

We first show that the first-order self-energy, represented by the diagram in Fig.~\ref{fig:sigma}(a), vanishes. It is straightforward to see that the first-order self-energy is independent of $\bm k$ and $\omega$, and does not have an imaginary part. Hence, it represents a constant energy shift of the $f$ electron, which must vanish because of the particle-hole symmetry in the model \eqref{eq:H}.

With the previous discussions, it becomes clear that only $f$ electron has a nontrivial self-energy, with corrections starting at second-order perturbations in $U$. In the rest of this section, we will focus on the second-order self-energy of the $f$ electrons.
Next, we argue that, due to the heaviness of the $f$ band, in the limit that the temperature $T$ is much larger than the width of the $f$ band $\Lambda_f$, the self-energy is approximately momentum-independent.

\begin{figure}[htbp]
\includegraphics{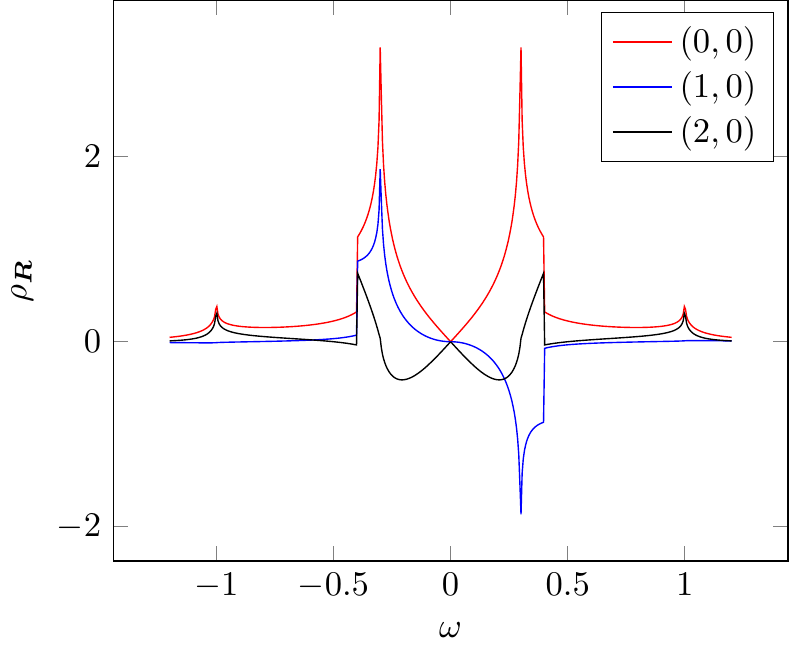}
\caption{Plot of the $f$-electron density of state $\rho_R(\omega)$ defined in Eq.~\eqref{eq:rho}. The red, blue and black lines represent the spatial components $R=(0, 0)$, $(1, 0)$ and $(2, 0)$, respectively.
The density of state is computed with parameters $t=1$, $v=0.5$ and $t^f=0.1$.}
\label{fig:rho}
\end{figure}

To proceed, it is convenient to evaluate the Feynman diagram in Fig.~\ref{fig:sigma}(b) in real space. The real-space self-energy $\Sigma(\bm r, \omega)$ has the form~\cite{Schweitzer1989,Schweitzer1990}
\begin{equation}
  \label{eq:Sigmar}
  \Sigma_{\bm x}(\omega)=\int_{-\infty}^\infty d\omega_1d\omega_2d\omega_3
  \rho_{\bm x}(\omega_1)\rho_{\bm x}(\omega_2)\rho_{\bm x}(\omega_3)
  \frac{n_F(\omega_1)n_F(\omega_2)n_F(-\omega_3)+
    n_F(-\omega_1)n_F(-\omega_2)n_F(\omega_3)}
  {\omega-\omega_1-\omega_2+\omega_3+i0^+},
\end{equation}
where $n_F(\omega)=1/(e^{\beta\omega}+1)$ is the Fermi-Dirac distribution function,
and $\rho_{\bm x}(\omega)$ is the $f$-electron density of state with the displacement $\bm x$, which can be read out from the bare Green’s function $G_0(\bm k,\omega)=[\omega = H_0(\bm k)]^{-1}$ using the expression of $H_0$ in \eqref{eq:H0},
\begin{equation}
  \label{eq:rho}
  \rho_{\bm x}(\omega) = -\frac1\pi\sum_k \im G^{0, ff}(\bm k, \omega)e^{i\bm k\cdot\bm x}.
\end{equation}
the $\bm x=0$ component, $\rho_0(\omega)$, represents the local density of state of the $f$ electron.
In the Anderson lattice model \eqref{eq:H}, the $f$ electron has a narrow band, as $t^f\ll t$, which is further broadened by the hybridization with the $c$ electrons. Therefore, the band width of the $f$ electron, $\Lambda^f$, is of the order of $v$.
This is illustrated in Fig.~\ref{fig:rho}, where $\rho_{\bm x}(\omega)$ is plotted.
As shown in this figure, the density of state vanishes linearly at $\omega = 0$, because the dispersion has a node at $\omega=0$; starting from $\omega =0$, the density of state increases until $\omega$ reaches a van Hove singularity. The density of state decreases pass that singularity, and further drops at another singularity. The majority of the density of state is within $-v<\omega<v$.

\begin{figure}
\subfigure[\label{fig:ims:a}]{\includegraphics[width=.45\textwidth]{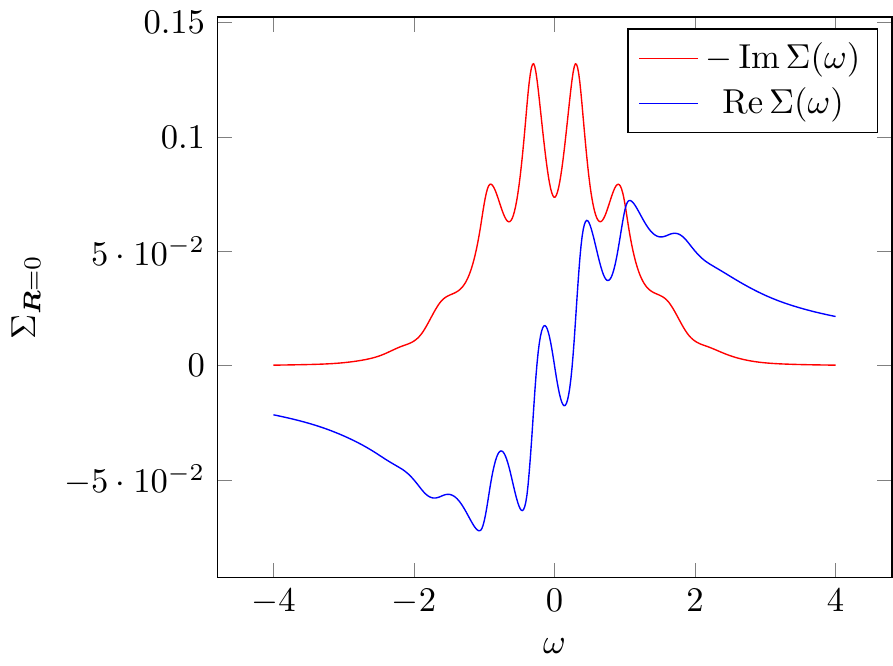}}
\subfigure[\label{fig:ims:b}]{\includegraphics[width=.45\textwidth]{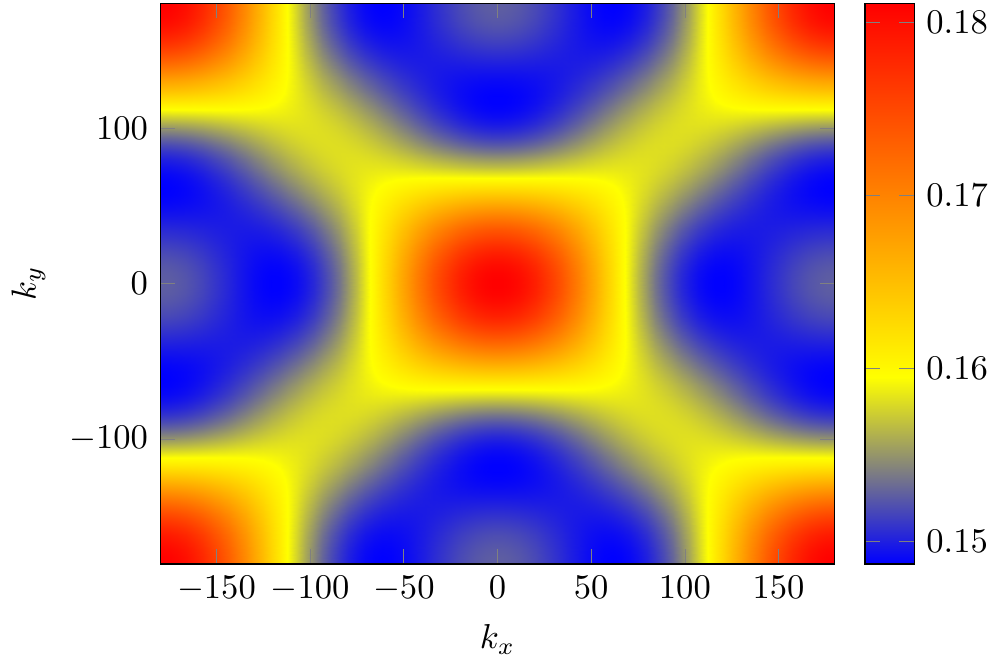}}
\caption{Numerical plots of the the self-energy computed from second-order perturbation theory of $U$. (a) The real and imaginary part of the local $\bm R=0$ component of the self energy. (b) The momentum dependence of the imaginary part of the zero-frequency self energy $\im\Sigma(\bm k, \omega=0)$, computed by Fourier-transforming $\Sigma_{\bm R}(\omega)$ with the spatial components $\bm R=(0, 0)$, $(1, 0)$, $(0, 1)$, $(1, 1)$, $(2, 0)$ and $(0, 2)$. The plot is produced with parameters $t=1$, $v=0.5$, $t^f=0.1$ and $U=1$.}
\label{fig:ims}
\end{figure}

Now, we consider the limit $T\gg\Lambda_f= v$.
In this limit, $\Sigma_{\bm x}$ is only nonvanishing at $\bm x=0$.
In other words, the Fourier transform, $\Sigma(\bm k, \omega)$, is momentum independent.
This is because the momentum dependence of $\Sigma(\bm k, \omega)$ comes from the dispersion of $f$ electrons, as $\Sigma$ is the convolution of three $f$-electron density-of-state functions in Eq.~\eqref{eq:Sigmar}.
In the limit $T\gg\Lambda_f= v$, the finite-temperature broadening is much larger than the bandwidth of $f$-electron, and therefore smears out the momentum dependence in $\Sigma(\bm k,\omega)$.
The fact that $\Sigma(\bm k, \omega)$ is approximately momentum independent is confirmed by our numerical simulation: $\im\Sigma(\omega=0)$ plotted in Fig.~\ref{fig:ims}(b) indeed shows very weak momentum dependence. Therefore, in this paper, we ignore the momentum dependence of $\Sigma$, and approximate $\Sigma(\bm k, \omega)=\Sigma(\omega)=\Sigma_0(\omega)$.

Next, we discuss the frequency dependence of the self energy $\Sigma_{\bm R=0}(\omega)$.
Since it is a convolution of three $\rho_0(\omega)$, which has a bandwidth of $\Lambda^f\sim v$, $\im\Sigma$ has a peak with width at the order of $3v$.
Indeed, the numerical results of $\im\Sigma(\omega)$, shown in Fig.~\ref{fig:ims}(b), shows a broad peak with a width on the order of $3v$.
Besides the overall features at the scale of $\omega\sim3v$, $\Gamma(\omega)$, shown in Fig.~\ref{fig:ims}(a), also exhibits oscillations, with the first peak appearing at the energy corresponding to the first van Hove singularity that appeared in the density of state shown in Fig.~\ref{fig:rho}.
This oscillation is an echo of the van Hove singularity, broadened because three $\rho(\omega)$ functions are convoluted.
In particular, focusing on $\omega$ close to zero, $\Gamma(\omega)$ increases with $\omega$.
This can be understood by recalling that $\rho(\omega)$ vanishes at $\omega=0$, and increases as $\omega$ increases.
Since $\Gamma(\omega)$ is even in $\omega$, it can be approximated as
\begin{equation}
\label{eq:ims=a+bw2}
\Gamma(\omega)=-\im\Sigma(\omega)=\gamma_0+\gamma_2\omega^2,\quad a,c>0.
\end{equation}
Correspondingly, $\re\Sigma(\omega)$ decreases with $\omega$ for small omega. As an odd function of $\omega$, it can be approximated by the following form for small $\omega$,
\begin{equation}
\label{eq:res=bw}
\re\Sigma(\omega) = -\gamma_1\omega,\quad b>0.
\end{equation}

Combining Eqs.~\eqref{eq:ims=a+bw2} and \eqref{eq:res=bw} together, we obtain the following small-frequency expansion,
\begin{equation}
\label{eq:ssw}
\Sigma = -i\gamma_0 - \gamma_1\omega - i\gamma_2\omega^2.
\end{equation}
Here, the positivity of the coefficients $\gamma_{0,1,2}$ is a consequence of the vanishing density of state at $\omega=0$. Therefore, we expect this to be a generic feature in models of Kondo insulators and semimetals with point nodes, both of which have $\rho_0(\omega)=0$.

\vspace{0.5cm}

\begin{figure}
\begin{center}
    \begin{tabular}{p{0.8 \columnwidth}} 
     \resizebox{0.8 \columnwidth}{!}{\includegraphics{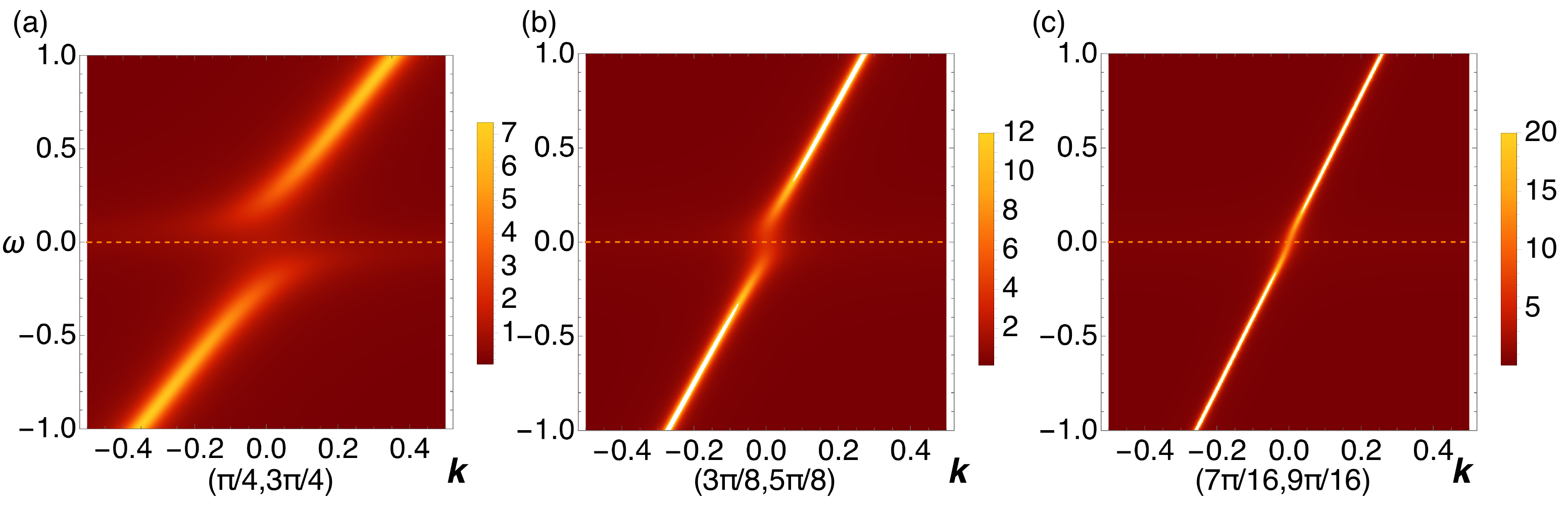}}
   \end{tabular}
\end{center}
\caption{
\label{fig:fig1sup}
(Color online)
Spectral functions defined in Eq.~(\ref{eq:spec}) as a function of $\omega$ and $\Vec{k}$ on the line cuts defined in Fig.~\ref{fig:fig2}(a) at $T=1/60$. We set $U = 8$.
}
\end{figure}

\begin{flushleft}
{\bf S4. Spectral functions as a function of $\omega$ and $\Vec{k}$}
\end{flushleft}
We show the spectral functions as a function of $\omega$ and $\Vec{k}$ obtained from the DMFT calculation, whose parameters are same as that in Fig.\ref{fig:fig1sup}.
The line cuts defined in Fig.~\ref{fig:fig2}(a) at $T=1/60$.

\begin{flushleft}
{\bf S5. Temperature and hybridization dependence of the lifetime}
\end{flushleft}

\begin{figure}
\begin{center}
    \begin{tabular}{p{0.4 \columnwidth}} 
     \resizebox{0.4 \columnwidth}{!}{\includegraphics{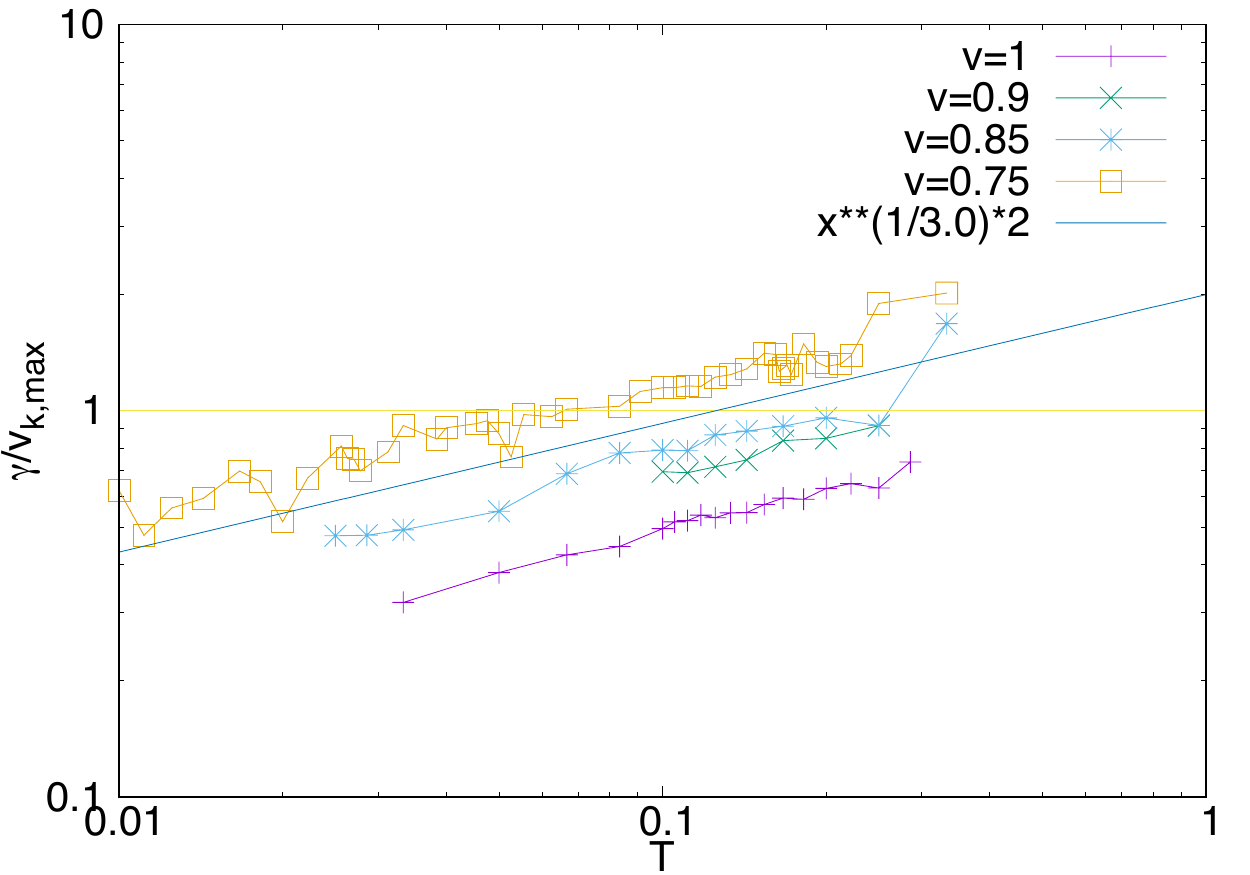}}
   \end{tabular}
\end{center}
\caption{
\label{fig:fig2sup}
(Color online)
Temperature dependence of the lifetime with different hybridization strength in the two-dimensional system with $d$-wave hybridization $V_{\Vec{k}}=v (\cos k_{x} - \cos k_{y}) \sigma_{0}$. The solid line indicates a guide.
The exceptional points and Fermi arcs in the momentum space appear when $\gamma/v_{\Vec{k},{\rm max}} < 1$ is satisfied.
At different hybridization strengths, the Fermi arcs appear at different temperatures.
}
\end{figure}

We show the temperature and hybridization dependence of the lifetime.
The inverse lifetime is defined as $\gamma = -{\rm Im}\Sigma_{f}(\omega=0)/2$.
When $\gamma/v_{\Vec{k},{\rm max}} < 1$ is satisfied, the exceptional points in the momentum space appear as shown in Fig.~4(a) in the main text.
Here, $v_{\Vec{k},{\rm max}} = 2v$ is a maximum value of the hybridization $v_{\Vec{k}}=v(\cos k_{x}-\cos k_{y})$ in the momentum space.
Figure \ref{fig:fig2sup} shows that, at different hybridization strengths, the Fermi arcs appear at different temperatures.

\end{document}